\documentclass[twocolumn,aps,prl,showpacs,floatfix]{revtex4}
\usepackage{graphicx}
\usepackage{times}
\usepackage{nicefrac}
\usepackage{amsmath}
\usepackage{amsfonts}
\usepackage{amssymb}
\usepackage{amsthm}
\usepackage{epsf}
\usepackage{bm}
\usepackage{times}

\usepackage{dcolumn}

\newcolumntype{.}{D{x}{}{-1}}

\newcommand{\balpha}{{\mbox{\boldmath$\alpha$}}}
\newcommand{\bnabla}{{\mbox{\boldmath$\nabla$}}}

\newcommand{\be}{\begin{eqnarray}}
\newcommand{\ee}{\end{eqnarray}}

\usepackage{color}
\definecolor{BLUE}{rgb}{0.0,0.0,1.0}

\usepackage{soul}
\soulregister\cite7
\soulregister\ref7

\begin{document}

\title{QED corrections to the $^2P_{1/2}-{}^2P_{3/2}$ fine structure in fluorinelike ions:\\
  model Lamb shift operator approach}

\author{V. M. Shabaev$^1$, I. I. Tupitsyn$^1$, M. Y. Kaygorodov$^1$, Y. S. Kozhedub$^1$, A. V. Malyshev$^1$, and D. V. Mironova$^{2}$  }

\affiliation {
$^1$ Department of Physics, St. Petersburg State University, Universitetskaya 7/9, 199034 St. Petersburg, Russia \\
$^2$ St. Petersburg Electrotechnical University LETI, Prof. Popov Str. 5, 197376 St. Petersburg, Russia
}

\begin{abstract}

  In [Li {\it et al.,} Phys. Rev. A {\bf 98}, 020502(R) (2018)] it was claimed that the model-potential computations
  of the Lamb shift on the  $^2P_{1/2}-{}^2P_{3/2}$ fine structure in fluorinelike uranium lead to a discrepancy between
  theory and experiment. Later, it was reported by [Volotka {\it et al.}, Phys. Rev. A {\bf 100}, 010502(R) (2019)]
  that  {\it ab initio} QED calculation, including the first-order one-electron QED contributions and
  the related effects of two-electron  screening,  yields the result which
   restores the agreement between theory and experiment and  strongly disagrees
  with the model-potential Lamb shift values. In the present paper, the model Lamb shift operator 
  [Shabaev {\it et al.,} Phys. Rev. A {\bf 88}, 012513 (2013)]  
  is used
  to evaluate the QED effects on the  $^2P_{1/2}-{}^2P_{3/2}$ fine structure in F-like ions.
The calculations are performed by incorporating
this operator into the Dirac-Coulomb-Breit equation employing
different methods. It is demonstrated that the methods, based on including 
the Lamb shift operator either into the Dirac-Fock equations 
or into the calculations by perturbation theory,  lead to the theoretical results
which are in good agreement with each other and 
with experiment. The restriction of these results
to the first order in the QED effects leads to a value which
agrees with the aforementioned {\it ab initio} QED result.

\end{abstract}
\pacs{12.20.Ds}
\maketitle

\section{Introduction}

High-precision measurements with many-electron atoms and ions require accurate theoretical
calculations including relativistic, electron-correlation and quantum electrodynamics (QED)
effects. While the relativistic and electron-correlation contributions are generally
taken into account within the framework of the Dirac-Coulomb-Breit (DCB) Hamiltonian,
 {\it ab initio} evaluation of the QED corrections requires the use of  perturbation theory (PT) methods.
For middle- and high-$Z$ few-electron ions, the PT calculations start with the Dirac
equation for an electron moving in the Coulomb field of the nucleus and include
the QED corrections up to the first or second order in the parameter $1/Z$ (see, e.g., Refs.~\cite{sha18,ind19}
and references therein). The same method can also be applied for many-electron
atoms, provided the PT starts with an effective potential, 
which partially takes into account
the electron-electron interaction effects~\cite{sap02,sha05,che06,art07,yer07,mal17}.

Since  {\it ab initio} QED calculations are rather complicated even in the lowest-order case,
there exists
a great demand in some approximate methods which could allow one to easily incorporate the
QED corrections into the calculations based on the DCB Hamiltonian. To this end, a number
of such methods has been proposed \cite{ind90,pyy03,dra03,fla05,thi10,pyy12,tup13,sha13,sha15,tup16,gin16}.
All these methods exploit the idea of scaling  
the Lamb shift results obtained for the Coulomb potential to
other atomic potentials, which take into account the screening
effects. In Ref. \cite{sha13}, to perform such a scaling, first, the diagonal
and nondiagonal matrix elements of the lowest-order one-electron QED corrections for the case
of the Coulomb potential
have been calculated in a wide range of the nuclear charge number $Z$.
Then, these results have been  used to model the Lamb shift 
operator by a sum of local and nonlocal potentials with the parameters
of these potentials fitted to the diagonal and nondiagonal Lamb shift matrix elements in
the Coulomb field. The obtained model Lamb shift (MQED) potential can be easily included into 
calculations based on the DCB Hamiltonian. This can be done in various ways, from
evaluating 
the contribution  of this potential by perturbation theory
to including this potential into the Dirac-Fock (DF) or the configuration-interaction
Dirac-Fock (CI-DF) equations \cite{sha13,sha15,tup16}.

The MQED operator has been successfully applied to calculations of the QED corrections
to the binding energies in various atomic systems \cite{sha13,sha15,tup16,pas17,yer17,mac18,si18,mue18,kay19}.
However, in Ref. \cite{li18}
it was claimed that the evaluation of the QED corrections to the 
$^2P_{1/2}-{}^2P_{3/2}$ fine structure in F-like uranium, based on different
approximate Lamb shift potentials (including the  
MQED operator suggested in Ref. \cite{sha13}),
leads to a discrepancy between theory and experiment.
In Ref.  \cite{vol19}, it has been reported that
 {\it ab initio} QED calculation, which includes the first-order one-electron QED
contributions and the  screened self-energy and vacuum-polarization corrections, restores
the agreement with experiment.
 Again, this paper contains a statement that the
{\it ab initio} QED results strongly disagree with the model-potential values.
While in the case of F-like ions the {\it ab initio} QED  calculations can be performed
with the help of the same methods as for Li- and B-like ions (see, e.g., Ref. \cite{mal17}
and references therein), the corresponding calculations for numerous other systems are
 much more difficult. This concerns, e.g., atoms and ions with complex electronic
 structure \cite{tup16} as well as autoionizing states 
 \cite{Yerokhin:2017:042505:note, Yerokhin:2018:023105, zay19, zay19_os}.
 Therefore, it is extremely important to have simple model-potential
 methods
which could allow one to account for the QED corrections to a reasonable accuracy. 
To this end, in the present work the aforementioned statements of Refs. \cite{li18,vol19} 
are examined by calculations of the QED corrections to the
$^2P_{1/2}-{}^2P_{3/2}$ fine structure in F-like ions, using the MQED operator.
The calculations are performed by means of five different methods: (1)~evaluation of the
expectation value of the MQED operator with the DF wave function,
 (2)~evaluation of the
expectation value of the MQED operator with the CI-DF wave function,
(3)~including
the  MQED operator into the DF equations self-consistently,
(4)~including the MQED operator into the CI-DF Hamiltonian self-consistently,
and (5)~including the MQED operator into the PT
calculations up to the second order in $1/Z$.
We find that
the last three methods yield the results which, being very close to each other, 
are in good agreement with experiment for fluorinelike uranium. The restriction  of these results
to the first order in the QED effects gives  a value which is
close to the {\it ab initio} result of Ref. \cite{vol19}.

The relativistic units ($\hbar=c=1$) are used throughout the paper.

%

\section{Model Lamb shift operator}

The one-electron Lamb shift operator can be approximated by a sum of
the self-energy (SE) and vacuum-polarization (VP) operators,
\begin{eqnarray}
h^{\rm QED} = h^{\rm  SE}+ V^{\rm VP}\,.
\end{eqnarray}  
The local vacuum-polarization potential is given, in turn, by a sum of the Uehling
and Wichmann-Kroll potentials, $ V^{\rm VP} =V^{\rm Uehl} + V^{\rm WK}$.
The direct calculation of the Uehling potential, which gives the dominant contribution
to   $ V^{\rm VP}$,
causes no problem.  To a good accuracy, it can also be evaluated by the use of approximate
formulas from Ref. \cite{ful76}. The evaluation of the Wichmann-Kroll potential is a much more
complicated task. However, since this term is generally much smaller than the Uehling one,
 it can be evaluated for the pointlike nucleus  with the help of approximate formulas
 from Ref. \cite{fai91}. These calculation methods have been incorporated into
 the Fortran package QEDMOD presented in Ref. \cite{sha15}.

 In the model Lamb shift operator approach \cite{sha13}, the self-energy
 operator is represented as a sum of local and nonlocal parts,
\begin{eqnarray}
  h^{\rm SE} = h^{\rm  SE}_{\rm loc}+ h^{\rm SE}_{\rm nl}\,,
\end{eqnarray} 
where $h^{\rm  SE}_{\rm loc}$ is actually a semilocal operator, acting differently on wave functions of different angular symmetry, and $ h^{\rm SE}_{\rm nl} $ is a nonlocal operator.
For a given angular Dirac symmetry $\kappa=(-1)^{j+l+1/2}(j+1/2) $,
the semilocal part is defined by
\begin{eqnarray}
 h^{\rm  SE}_{\rm loc,\kappa} =A_{\kappa} \exp{(-r/\lambda_{\rm C})}\,,
\end{eqnarray} 
where $\lambda_{\rm C}=\hbar/(mc)$ and the constant $ A_{\kappa} $ is determined from the condition that the matrix element
of $ h^{\rm  SE}_{\rm loc,\kappa}$  calculated with the hydrogenlike wave function of the lowest
energy state for the given $\kappa$ reproduces the exact value of the SE shift.
The nonlocal operator is given in a separable form,
\begin{eqnarray}
h^{\rm SE}_{\rm nl}  = \sum_{i,k=1}^{n} |\phi_i\rangle B_{ik}\langle \phi_k|\,.
\end{eqnarray} 
The functions $\phi_i $ play a role of the projector functions. The choice of
these functions was described in detail in Ref. \cite{sha13}.
The constants $B_{ik}$ are determined by the condition that the diagonal and nondiagonal matrix
elements of $ h^{\rm SE} $ calculated with hydrogenlike wave functions $\psi_i $ are equal to the
exact values of the one-loop self-energy contributions \cite{sha93},
\begin{eqnarray}
  \langle \psi_i| h^{\rm SE}|\psi_k\rangle  = \frac{1}{2}
      \langle \psi_i|[\Sigma(\varepsilon_i) + \Sigma(\varepsilon_k)]|\psi_k\rangle \,.
\end{eqnarray} 
Then, one obtains
\begin{eqnarray}
  B_{ik}& =& \sum_{j,l=1}^{n} (D^{-1})_{ji}\langle \psi_j|\{\frac{1}{2}[\Sigma(\varepsilon_i) 
    + \Sigma(\varepsilon_k)]\nonumber\\
    && - h^{\rm  SE}_{\rm loc}\}|\psi_l\rangle (D^{-1})_{lk}\,,
\end{eqnarray} 
where $D_{ik}=\langle \phi_i|\psi_k\rangle $. The computation code based on this method
was published in Ref. \cite{sha15}.

The total model Lamb shift operator for a many-electron atom is given by
\begin{eqnarray} \label{mqed}
H^{\rm QED}=\sum_{i} h^{\rm QED}_i\,,
\end{eqnarray}
where the summation runs over all atomic electrons.

\section{Calculations and results}

Instead of {\it ab initio} QED calculations, we incorporate the model Lamb
shift operator into
the calculations based on the Dirac-Coulomb-Breit Hamiltonian.
The standard form of the DCB Hamiltonian is given by:
\begin{eqnarray} \label{dcb}
H^{\rm DCB}=
\Lambda^{(+)}\Bigl[\sum_{i}h_i^{\rm D}  +\sum_{i<k}V_{ik}\Bigr]\Lambda^{(+)}\,,
\end{eqnarray}
where the indices $i$ and $k$ enumerate the atomic electrons,
$h_i^{\rm D}$ is the one-electron Dirac Hamiltonian, and
\begin{eqnarray} \label{br}
  V_{ik} &=& V^{\rm C}_{ik}+ V^{\rm B}_{ik}\nonumber \\
&=&\frac{\displaystyle \alpha}{\displaystyle r_{ik}} 
-\alpha\Bigl[\frac{\displaystyle{ {\balpha}_i\cdot {\balpha}_k}}
{\displaystyle{ r_{ik}}}+\frac{\displaystyle 1}{\displaystyle 2}
( {\balpha}_i\cdot{\bnabla}_i)({ {\balpha}_k\cdot\bnabla}_k)
r_{ik}\Bigr]
\,
\end{eqnarray}
is the electron-electron interaction operator within the Breit approximation.
The operator $\Lambda^{(+)}$ is the projector on the states constructed from the positive-energy
eigenfunctions of some one-particle Dirac Hamiltonian $\tilde{h}^{\rm D}$. 
The role of $\tilde{h}^{\rm D}$ can be played, e.g., by the Dirac Hamiltonian 
with the Coulomb or an effective potential or the nonlocal DF operator~$h^{\rm DF}$.

In the present work, to find the eigenvalues and eigenfunctions of the DCB Hamiltonian,
we use the configuration-interaction Dirac-Fock-Sturm (CI-DFS) method \cite{tup03,tup18}. 
The many-electron wave function $\psi(\gamma J)$, with $J$ being
the total angular momentum and $\gamma$ standing for all other quantum
numbers, is expanded in terms of a large number of the configuration-state functions (CSFs):
\begin{equation}
\label{psi}
\psi(\gamma J)=\sum_{\alpha}c_{\alpha} \Phi_{\alpha}(J),
\end{equation}
where $\Phi_{\alpha}(J)$, being the eigenfunctions of the square of total
angular momentum $J^2$, correspond to a given relativistic configuration.
They are obtained as linear combinations of the Slater determinants.
The one-electron orbitals corresponding to the occupied shells  ($\phi_j$)
are obtained from the DF equations, while the vacant orbitals  ($\tilde{\phi}_j$)
are determined by solving the Dirac-Fock-Sturm equations,
\begin{equation}
  (h^{\rm DF}-\varepsilon_{j_0})\tilde{\phi}_j=\lambda_j W(r)\tilde{\phi}_j\,,
\end{equation}  
where $\varepsilon_{j_0}$ is the one-electron energy of an occupied DF
orbital and $W(r)$  is a constant sign weight function.
The parameter $\lambda_j$ is defined as an  eigenvalue of
the Sturmian operator. The  weight function $W(r)$
is taken to be
\begin{equation}
W(r)=\frac{1-\exp{[-(\mu r)^2]}}{(\mu  r)^2} \,.
\end{equation} 
With this choice, it  is regular at the origin and goes to zero like $1/r^2 $
at $r\rightarrow \infty$. For $\lambda_j =0$ the Sturmian function coincides with
the reference DF orbital  ($\phi_{j_0}$). All the Sturmian functions have the same
exponential asymptotics at  $r\rightarrow \infty$ as the reference DF wave function. 
Since the Sturmian operator is Hermitian
and does not contain continuum spectra, the Sturmian eigenfunctions 
form a discrete and complete basis set of one-electron wave functions.

\begin{table}
\renewcommand{\arraystretch}{1.15}
\begin{center}
\caption{QED contributions to the $^2P_{1/2}$ and $^2P_{3/2}$ energy levels and their difference in F-like ions, in eV.
  The DF$_{\rm av}$ and  CI-DFS$_{\rm av}$ values are obtained by averaging the model Lamb shift operator
  with the DF and CI-DFS wave functions, respectively. The vacuum-polarization contribution consists of
   the Uehling and Wichmann-Kroll (WK) terms. 
}
\label{tab1}
%
\resizebox{\columnwidth}{!}{%
\begin{tabular}{lcrrrr}
  \hline

$Z=42$ & & & & & \\
\hline
  Method & State &  SE \hspace{2mm} 
  &  Uehling  
 &  WK  \hspace{1mm} 
&   QED$_{\rm tot}$    \\
\hline
  DF$_{\rm av}$   & $^2P_{1/2}$  & 48.2040 & $-$5.5529 &    0.0846 &  42.7357 \\
                  & $^2P_{3/2}$  & 47.9671 & $-$5.5575 &    0.0847 &  42.4943 \\
    &  $^2P_{1/2}- {}^2P_{3/2}$  &  0.2369 &    0.0046 & $-$0.0001 &   0.2414 \\
\hline
CI-DFS$_{\rm av}$ & $^2P_{1/2}$  & 48.0982 & $-$5.5418 &    0.0844 &  42.6408 \\
                  & $^2P_{3/2}$  & 47.8610 & $-$5.5464 &    0.0845 &  42.3991 \\
    &  $^2P_{1/2}- {}^2P_{3/2}$  &  0.2372 &    0.0046 & $-$0.0001 &   0.2417 \\
\hline

$Z=92$ &&&&& \\
\hline
  Method & State &  SE    \hspace{2mm} 
  &  Uehling  
 &  WK  \hspace{1mm} 
&   QED$_{\rm tot}$    \\
\hline

  DF$_{\rm av}$   & $^2P_{1/2}$  & 858.7834 & $-$217.7821 &   12.3823 &  653.3836 \\
                  & $^2P_{3/2}$  & 858.2538 & $-$219.9653 &   12.5367 &  650.8252 \\
    &  $^2P_{1/2}- {}^2P_{3/2}$  &   0.5296 &      2.1832 & $-$0.1544 &    2.5584 \\
\hline
CI-DFS$_{\rm av}$ & $^2P_{1/2}$  & 855.5099 & $-$217.0007 &   12.3405 &  650.8497 \\
                  & $^2P_{3/2}$  & 854.8638 & $-$219.1415 &   12.4920 &  648.2144 \\
    &  $^2P_{1/2}- {}^2P_{3/2}$  &   0.6460 &      2.1408 & $-$0.1515 &    2.6353 \\

\hline

\end{tabular}%
}
\end{center}
\end{table}

To evaluate the Lamb shift, we use different methods. As the first step, we have
evaluated the average values of the MQED operator with the DF and the CI-DFS wave functions.
The related SE and VP contributions for the $^2P_{1/2}$ and  $^2P_{3/2}$ states of F-like ions
with $Z=42, 92$  are given in Table \ref{tab1}.  The DF and CI-DFS results are labeled as 
DF$_{\rm av}$ and CI-DFS$_{\rm av}$, respectively.
It can be seen that there exists a strong cancellation
of the QED corrections in the  $^2P_{1/2}-{}^2P_{3/2}$ transition energy.
Due to this cancellation, in case of  $Z=92$ the SE contribution is even smaller than
the VP contribution.
This means that for the transition under consideration
the QED contribution is very sensitive to the inter-electron interaction
and, therefore, the MQED operator should be incorporated into the calculations 
in a more comprehensive way. To this end, in addition to averaging
the MQED operator with the DF and CI-DFS wave functions,  we have performed
the calculations by means of three more elaborate methods.
The corresponding results are presented in Table \ref{tab2}.
The DF$_{\rm scf}$ indicates the results obtained by including the MQED operator into the DF equations
self-consistently.
The  CI-DFS$_{\rm scf}$
value presents the results obtained by the CI-DFS method with the MQED operator incorporated into
the DF and CI-DFS equations. 
The latter implies the related modification of the projector operators~$\Lambda^{(+)}$ as well.
Finally, the PT$_{\rm scf}$ value denotes the QED corrections obtained
by calculations of the binding energies by the PT up to the second order 
in $1/Z$ employing the basis of hydrogenlike wave functions with the MQED operator
included into the Dirac Hamiltonian. It should be noted that we have
also performed the PT calculations starting from the DF Hamiltonian as
the zeroth-order approximation and employing the corresponding DF basis instead of the H-like basis.
The values which are in perfect agreement with the DF$_{\rm scf}$ and CI-DFS$_{\rm scf}$ results have
been obtained in this case.

\begin{table}
\renewcommand{\arraystretch}{1.2}
\begin{center}
\caption { QED contributions to the $^2P_{1/2}-{}^2P_{3/2}$ transition energy in F-like ions, in eV.
The  DF$_{\rm av}$ and  CI-DFS$_{\rm av}$ values from Table~\ref{tab1} are compared with the more elaborate
calculations: DF$_{\rm scf}$ indicates the result obtained by including the MQED operator into the DF equations
self-consistently, CI-DFS$_{\rm scf}$ value presents the result obtained by the CI-DFS method with the MQED operator 
incorporated into the DF and CI-DFS equations, and PT$_{\rm scf}$ value denotes the QED correction obtained
by calculations of the binding energy by the perturbation theory up to the second order 
in $1/Z$ on the basis of hydrogenlike wave functions with the MQED operator
included into the Dirac Hamiltonian. }
\label{tab2}
\vspace{1mm}
\begin{tabular}{c@{\quad}lllll}
\hline 
%
$Z$ &  \multicolumn{1}{c}{DF$_{\rm av}$~~}   
&      \multicolumn{1}{c}{CI-DFS$_{\rm av}$~~}
&      \multicolumn{1}{c}{DF$_{\rm scf}$~~}  
&      \multicolumn{1}{c}{CI-DFS$_{\rm scf}$~~}  
&      \multicolumn{1}{c}{PT$_{\rm scf}$~~}  \\
\hline
%
42 &    0.241  &  0.242  &  0.238  &  0.238   &    0.239    \\
92 &    2.56   &  2.64   &  2.12   &  2.12    &    2.10     \\
\hline

\end{tabular}
\end{center}
\end{table}

The difference between the  CI-DFS$_{\rm av}$ 
results, from one side, and the  DF$_{\rm scf}$, CI-DFS$_{\rm scf}$, and  PT$_{\rm scf}$  results,
from the other side,
is due to the
single-particle excitation into the negative-energy continuum
and the higher-order QED effects.
Both effects are automatically included in
the last three methods but not included in the  CI-DFS$_{\rm av}$ method. To check this,
we have performed the calculations of the negative-energy contribution and the second-order
QED effect separately.

First, to study the importance of the second-order QED effect,
we have calculated the first- and second-order QED contributions
by the  CI-DFS$_{\rm scf}$ method. This has been done by introducing a parameter
$\lambda$ in front of the MQED operator in the DF and CI-DFS equations
and representing the total energy as
\begin{eqnarray}\label{expan}
  E(\lambda)=E(0)+\frac{dE}{d\lambda}\Bigr|_{\lambda=0}\lambda+
  \frac{1}{2} \frac{d^2E}{d\lambda^2}\Bigr|_{\lambda=0}\lambda^2+\cdots\,.
\end{eqnarray}
Then, the first- and second-order QED contributions are given by the corresponding
expansion coefficients in Eq.~(\ref{expan}). 
These contributions for F-like uranium are presented in Table~\ref{tab3}. 
Our first-order QED
contribution,  $\frac{dE}{d\lambda}\Bigr|_{\lambda=0}$, should correspond 
to the {\it ab initio} calculation of Ref.~\cite{vol19}, which includes
the first-order self-energy and vacuum-polarization corrections and the related effects
of two-electron screening.
Indeed, Table~\ref{tab3} shows that our first-order
contribution to the fine-structure splitting
amounts to 2.34 eV, which is close to the
 {\it ab initio} contribution 2.47(2) eV of Ref.~\cite{vol19}.
In contrast to that, our total MQED values obtained by  DF$_{\rm scf}$,  CI-DFS$_{\rm scf}$,
and  PT$_{\rm scf}$ methods include also partly the higher-order QED effects.
As one can see from Table~\ref{tab3}, in case of F-like U the second-order QED contribution
amounts to $-$0.24 eV and  shifts the
QED correction to 2.10 eV. These results clearly demonstrate
the importance of the
higher-order QED contribution in the calculations of the fine-structure splitting,
which has been omitted in Ref. \cite{vol19}.

\begin{table}
\renewcommand{\arraystretch}{1.25}
\begin{center}
\caption{QED contributions to the $^2P_{1/2}$ and $^2P_{3/2}$ energy levels and their difference in F-like U, in eV.
The first- and second-order QED contributions are  obtained by the  CI-DFS$_{\rm scf}$ method as the first and
second derivatives with respect to the parameter $\lambda$ introduced in front of the MQED operator in the DF and CI-DFS equations. The derivatives are evaluated at $\lambda =0$. The sum of these terms (``Sum'') is compared with the total QED value (CI-DFS$_{\rm scf}$). }
\label{tab3}
\vspace{1mm}
\begin{tabular}{crrrr}
  \hline

\hline
State & $\frac{dE}{d\lambda}\Bigr|_{\lambda=0}$ &
 $\frac{1}{2}\frac{d^2E}{d\lambda^2}\Bigr|_{\lambda=0}$
  & Sum$\;\;$ & CI-DFS$_{\rm scf}$   \\
\hline

  $^2P_{1/2}$                & 652.47$\;$ &     1.18 $\;\;\;$ & 653.65 &  653.34$\;$ \\
  $^2P_{3/2}$                & 650.13$\;$ &     1.42 $\;\;\;$ & 651.55 &  651.22$\;$ \\
  $^2P_{1/2}- {}^2P_{3/2}$   &   2.34$\;$ &  $-$0.24 $\;\;\;$ &   2.10 &    2.12$\;$ \\

\hline

\end{tabular}
\end{center}
\end{table}

To examine the role of the negative-energy contribution, we have also evaluated this effect separately
by summing the single-particle excitation into the negative-energy continuum,
\begin{eqnarray} \label{neg}
  \Delta E _{\rm neg}^{\rm QED}
  &=& 2 \, \sum_{\varepsilon_p>0,\varepsilon_n<0} \,
\frac{\langle p \mid h^{\rm QED} \mid n \rangle}
     {\varepsilon_p \,-\, \varepsilon_n} \nonumber\\
     &&\times \langle \hat a^{+}_n \, \hat a_p \,\Psi \mid \hat H^{\rm DCB} \mid \Psi \rangle \, ,
\end{eqnarray}
where in the second matrix element we use the second-quantization picture with  $\hat a^{+}_n$ and  $\hat a_p$ being
the creation and annihilation operators for the negative- and positive-energy
states, respectively, and  $\varepsilon_n$ and $\varepsilon_p$ denote the corresponding one-electron
energies. The many-electron wave function $\Psi$ is assumed to be the solution of the DCB equation without the inclusion of the MQED operator.  Table \ref{tab4} presents
the results of these calculations for F-like U, together with the related
CI expectation values, $\Delta E^{\rm QED}_{\rm av}$, and the second-order QED contributions,  $\Delta E^{\rm QED}_{\rm s.o.}$. 
The calculations of all the contributions have been performed using the DF basis as well as the
H-like basis.
It can be seen that in both calculations the sum of the expectation value
and the negative-energy and the second-order QED contributions yields the results
which are in good agreement with each other and with the  CI-DFS$_{\rm scf}$ value
from Table \ref{tab2}.
Thus, the difference between the CI-DFS$_{\rm scf}$ result and the expectation CI-DFS$_{\rm av}$
value is indeed caused by the single-particle excitations into
the negative-energy continuum and  the higher-order QED effects
which give comparable contributions. It should be stressed that only
the first of these effects has been accounted in  Ref. \cite{vol19}.

\begin{table}
\renewcommand{\arraystretch}{1.2}
\begin{center}
\caption{QED contributions to the $^2P_{1/2}$ and $^2P_{3/2}$ energy levels and their difference in F-like U, in eV.
The calculations of the CI average value,   $\Delta E^{\rm QED}_{\rm av}$,   the negative-energy-continuum
contribution, $\Delta E^{\rm QED}_{\rm neg}$, which is defined by Eq.~(\ref{neg}),  
and the second-order QED effect, $\Delta E^{\rm QED}_{\rm s.o.}$, considered in Table~\ref{tab3}, 
are performed in two different basis. 
 The last column represents the sum of all the contributions.
  }
\label{tab4}
\vspace{1mm}
\begin{tabular}{lcrrrr}
  \hline

%
  \rule{0pt}{2.6ex}Basis & State &  $\Delta E^{\rm QED}_{\rm av}$ 
  &  $\Delta E^{\rm QED}_{\rm neg}$ 
    &  $\Delta E^{\rm QED}_{\rm s.o.}$
  &  $\Delta E^{\rm QED}_{\rm sum}$
  \\
\hline
 DF      &   $^2P_{1/2}$               & 650.85 &    1.61 &    1.18 &  653.64  \\
         &   $^2P_{3/2}$               & 648.21 &    1.91 &    1.42 &  651.54 \\
         &   $^2P_{1/2}- {}^2P_{3/2}$  &   2.64 & $-$0.30 & $-$0.24 &    2.10  \\
\hline
 H-like  &   $^2P_{1/2}$               & 651.33 &    1.12 &    1.18 &  653.63  \\
         &   $^2P_{3/2}$               & 648.53 &    1.59 &    1.42 &  651.54  \\
         &   $^2P_{1/2}- {}^2P_{3/2}$  &   2.80 & $-$0.47 & $-$0.24 &    2.09  \\
\hline

\end{tabular}
\end{center}
\end{table}

In Table \ref{tab5}, we compare our MQED results 
for the QED corrections to the
$^2P_{1/2}-{}^2P_{3/2}$ transition energy
obtained by  the CI-DFS$_{\rm scf}$ method (in the case under consideration
the DF$_{\rm scf}$ method yields the same values)
with the previous calculations from Refs. \cite{li18,vol19}.
In Ref. \cite{li18}, the QED corrections have been evaluated using the GRASP2K, Welton, and MQED operator
methods. The  GRASP2K is the QED correction from the original  GRASP2K calculations \cite{jon13},
while the Welton value
is based on Welton's concept of the SE contribution implemented according to Ref. \cite{low13}.
In fourth column of Table \ref{tab5}, we give  the MQED results taken from Refs. \cite{li18,vol19}.
In the case of F-like U, this value  is very different from all our MQED results
presented in Table \ref{tab2}. The origin of this difference is unclear to us.
The last column of Table \ref{tab5} presents
our MQED values obtained by the  CI-DFS$_{\rm scf}$ method. 
These values, which we consider as 
the most reliable results within the MQED operator approach, are generally
 in reasonable
 agreement with the {\it ab initio} result from Ref. \cite{vol19}, presented
 in the fifth column.
 It should be stressed again, however, that while both methods account for the first-order
 QED contributions (including the screened QED corrections), our MQED values include also the
 higher-order QED correction, which has been omitted in  Ref. \cite{vol19}.
 In the case of F-like U, the second-order QED correction, being equal to $-$0.24 eV, 
yields the major part
of the difference between our MQED value  and the {\it ab initio} first-order result. 
It exceeds by an order of magnitude the uncertainty indicated
 for the {\it ab initio} value in Ref. \cite{vol19}.

In Ref. \cite{vol19}, the experimental QED contributions have been derived by subtracting
the theoretical non-QED results \cite{li18}, which have been assumed to be
sufficiently accurate, from the related experimental values. The obtained
``experimental'' QED values, which include the experimental uncertainties only, 
have been compared with the approximate and {\it ab initio} QED calculations \cite{li18,vol19}.
Despite a good agreement between the experimental and the {\it ab initio} theoretical QED results,
obtained in Ref. \cite{vol19}, we do not think that this can be considered as a test of the
screened  QED effects (which in Ref. \cite{vol19} are termed as the many-electron
second-order  QED contributions).
First, we doubt that the uncertainty of the non-QED calculations \cite{li18} is small enough
to be ignored  in determining the experimental QED effect. 
Second, even if we admit a very high accuracy of
the non-QED calculations from Ref. \cite{li18}, to  test QED one should combine
the screened QED effect, evaluated in Ref. \cite{vol19}, with 
the QED contribution from the two-photon exchange diagrams, which has not yet been evaluated.
For comparison, in a similar case of the $^2P_{3/2} - {}^2P_{1/2}$ transition in B-like
uranium the QED part of the two-photon exchange evaluated starting from
 the core-Hartree (CH)  potential amounts to $-0.21$~eV, 
while the screened QED effect is 0.64 eV.
Finally, as noted above, due to a strong cancellation of the first-order QED corrections for the 
$^2P_{1/2}$  and   $^2P_{3/2}$ states, the higher-order QED corrections,
which are not included in the {\it ab initio} result of Ref. \cite{vol19},
become rather important and should be taken into account in the QED tests.
This is confirmed by the large value of the higher-order QED contribution in the case
of F-like U.
In view of the above, we restrict the comparison of the theoretical QED contribution
with  experiment for F-like uranium only \cite{bei98}. As noted in Ref. \cite{li18}, 
this data point causes no doubts in its experimental accuracy and clearly disagrees
with the theoretical predictions obtained by the use of
the approximate QED methods  in  Ref. \cite{li18}. In this case, the ``experimental'' QED contribution amounts
to 2.25(16) eV \cite{li18,vol19}. As one can see from Table \ref{tab2}, this value agrees with
our MQED values obtained by the  DF$_{\rm scf}$, CI-DFS$_{\rm scf}$, and PT$_{\rm scf}$ calculations.
This means that at present we have no reasons to doubt the ability of the MQED operator approach
for evaluating the QED corrections to the binding energies, provided the MQED operator
is incorporated in the DF equations self-consistently.
As to tests of QED beyond the lowest-order one-electron approximation,
all the QED corrections in the order under consideration, including  
the two-photon-exchange QED contributions, as well as the major higher-order QED corrections 
must be evaluated before any conclusions
can be made.

\begin{table}
\renewcommand{\arraystretch}{1.2}
\begin{center}
\caption { QED contributions to the $^2P_{1/2}-{}^2P_{3/2}$ transition energy in F-like ions, in eV.
  The  GRASP2K is the QED correction from the original  GRASP2K calculations \cite{jon13}, the Welton value
  is based on Welton's concept implemented according to Ref. \cite{low13}, and MQED
  stands for the calculations using the model Lamb shift operator \cite{sha13}. It should be noted that, in contrast to
  the {\it ab initio} calculation of Ref. \cite{vol19}, the MQED calculation in this work includes also the
  higher-order QED correction which in the case of F-like U amounts to about $-$0.24~eV and yields the major part
  of the difference between the present MQED value  and the {\it ab initio} result of  Ref. \cite{vol19}.
}
\label{tab5}
\vspace{1mm}
\begin{tabular}{clllll}
\hline 
%
\rule{0pt}{2.6ex}$\;\;\;\;\; Z \;\;\;\;\;$ & \hspace{0.1mm}  \hspace{-5mm} GRASP2K$^a$   \hspace{0.1mm}
&   \hspace{-1.5mm}  Welton$^a$    \hspace{0.1mm}
&   \hspace{-1.5mm}  MQED$^a$ \hspace{0.1mm}
&   \hspace{-1.5mm}  {\it Ab initio}$^b$ \hspace{0.1mm}
&   \hspace{-1.5mm}  MQED$^c$  \hspace{0.1mm} \\[0mm]
\hline
%
18 &    0.0049~ &  0.0049~ &  0.0063~  &    0.0055(7)  &  0.0064~ \\
22 &    0.0125~ &  0.0125~ &  0.0154~  &   0.0139(10)  &  0.0157~\\
26 &    0.0266~ &  0.0265~ &  0.0318    &  0.0292(16)  &  0.0326~ \\
28 &    0.0368~ &  0.0366~ &  0.0435   &  0.0404(16)  &  0.0447~ \\
36 &    0.108~  &  0.107~  &  0.123~    &   0.118(3)    &  0.128~  \\
39 &    0.150~  &  0.149~  &  0.171~   &   0.164(3)  &  0.177~  \\
40 &    0.167~  &  0.165~  &  0.189~    &    0.182(3)   &  0.196~  \\
42 &    0.203~  &  0.201~  &  0.229~   &    0.222(4) &  0.238~   \\
74 &    1.50~   &  1.48~   &  1.67~   &   1.78(1)    &  1.81~  \\
92 &    1.33~   &  1.48~   &  1.79~      &  2.47(2)    &  2.12~  \\
\hline
\multicolumn{6}{l}{\small $^a$ Ref. \cite{li18}}\\
\multicolumn{6}{l}{\small $^b$ Ref. \cite{vol19}}\\
\multicolumn{6}{l}{\small $^c$ This work }\\
\end{tabular}
\end{center}
\end{table}

\section{Conclusion}
We have examined the MQED operator approach for calculations of the
QED corrections to the $^2P_{1/2}-{}^2P_{3/2}$ transition energy  in F-like ions.
It has been found that, due to a strong cancellation of the QED contributions
in the expectation values of the MQED operator with the DF and CI-DF wave functions,
the calculations must be performed incorporating the MQED operator
into the DF equations self-consistently.
This approach allows one to take into account single-particle excitations
into the negative-energy continuum and to include partly the higher-order
QED corrections. It has been found that both these QED effects can be important.
The obtained results,
being restricted to the first order  in the QED effects
are 
in agreement with the {\it ab initio} QED calculations \cite{vol19}.
 In case of F-like uranium, our total QED results
 are in good agreement with the ``experimental'' QED
contribution which has been derived in Ref. \cite{vol19} using the experimental data from
Ref. \cite{bei98} and the non-QED contribution from Ref. \cite{li18}.
This clearly shows that at present there is no reasons to
 doubt the ability of the MQED operator method \cite{sha13,sha15}.
However,  to test QED beyond the lowest-order one-electron approximation
the evaluation of the two-photon exchange diagrams is needed.


\section{Acknowledgments}

V.M.S., M.Y.K., and A.V.M. acknowledge the support from the Foundation for the advancement of theoretical physics
and mathematics ``BASIS''.
This work was also supported by RFBR (Grant No. 18-03-01220) and by 
 SPSU-DFG (Grants No.11.65.41.2017 and No. STO 346/5-1).
I.I.T acknowledges also support from  SPbSU (COLLAB 2019: No. 37717909).
 The research was carried out using computational resources provided by Resource Center ``Computer Center of SPbSU''.
%
%



\end{document}